\documentclass[aps,prl,twocolumn,superscriptaddress,showpacs,amsmath,amssymb]{revtex4}
\usepackage{graphicx}
\usepackage[floatfix]{epsfig}
\usepackage{dcolumn}
\usepackage{bm}

\begin{document}

\def\etal{{\it et al.}}
\newcommand{\ltwid}{\mathrel{\raise.3ex\hbox{$<$\kern-.75em\lower1ex\hbox{$\sim$
}}}}
\newcommand{\gtwid}{\mathrel{\raise.3ex\hbox{$>$\kern-.75em\lower1ex\hbox{$\sim$
}}}}

\title{Pair Phase Fluctuations and the Pseudogap}

\author{T.~Eckl}
\affiliation{Institut f\"ur Theoretische Physik und Astrophysik,
Universit\"at W\"urzburg, Am Hubland, D-97074 W\"urzburg, Germany}
\author{D.~J.~Scalapino}
\affiliation{Department of Physics, University of California, Santa Barbara, CA 93106-9530 USA}
\author{E.~Arrigoni}
\affiliation{Institut f\"ur Theoretische Physik und Astrophysik,
Universit\"at W\"urzburg, Am Hubland, D-97074 W\"urzburg, Germany}
\author{W.~Hanke}
\affiliation{Institut f\"ur Theoretische Physik und Astrophysik,
Universit\"at W\"urzburg, Am Hubland, D-97074 W\"urzburg, Germany}

\date{\today}

\begin{abstract}
The single-particle density of states and the
tunneling conductance are studied for a two-dimensional BCS-like Hamiltonian with a
$d_{x^2-y^2}$-gap and phase fluctuations.  
The latter are treated by a classical
Monte Carlo simulation of an $XY$ model.
Comparison of our results
with recent scanning tunneling spectra of Bi-based high-$T_c$ cuprates supports the idea
that the pseudogap behavior observed in these experiments can be understood
as arising from phase fluctuations of a $d_{x^2-y^2}$ pairing gap whose
amplitude forms on an energy scale set by $T_c^{MF}$ well above
the actual superconducting transition.
\end{abstract}


\maketitle



Intensive research has focused on the pseudogap regime, which is observed in the high-$T_c$
cuprates below a characteristic temperature that is higher than the transition
temperature $T_c$. It occurs in a number of different experiments as a suppression of 
low-frequency spectral weight \cite{Loe96,Din96,Lor93,Tak91,Ore99,Ren98M,Kug01,Wan02}. 
This striking pseudogap behavior initiated a variety of proposals as to
its origin \cite{Cha01,SK95,Pin97,EK95,Ran92,Fra98,Kwo99,VTxx}, since the answer to this question may be
a key ingredient for the understanding of high-$T_c$ superconductivity.
At present, there is no agreement as to
which of these proposals is correct.
In part, this reflects the possibility that there may be
different pseudogap phenomena operating in different temperature and
doping regimes. In part, this is because of the difficulty in
determining the experimental consequences of the various
theoretical proposals.  In this paper, we focus on the pseudogap phenomena 
observed in scanning tunneling spectroscopy
measurements \cite{Ren98M,Kug01} on $Bi_2Sr_2CaCu_2O_{8+\delta}$
(Bi2212) and $Bi_2Sr_2CuO_{6+\delta}$ (Bi2201).
We provide a detailed numerical solution of a minimal
model which, however, contains the key ideas of the cuprate
phase fluctuation scenario: that is, we explore the notion
that the pseudogap observed in these experiments arises from phase
fluctuations of the gap \cite{Ran92,EK95,Fra98,Kwo99,Ren98M,Kug01}.  In this
scenario,
below a mean field temperature scale $T_c^{MF}$, a
$d_{x^2-y^2}$-wave gap amplitude is assumed to develop. However, the
superconducting transition is suppressed to a considerably lower
temperature $T_c$ by phase fluctuations \cite{EK95}.  In the 
intermediate temperature
regime between $T_c^{MF}$ and $T_c$, the phase fluctuations of the gap give
rise to pseudogap phenomena.

We will study as a model for phase fluctuations a two-dimensional
BCS Hamiltonian
\begin{equation}
H=-t \sum_{\langle i\,j\rangle,\sigma}(c^\dagger_{i\,\sigma}c_{j\,\sigma}+c^\dagger_{j\,\sigma}c_{i\,\sigma})-\frac{1}{4}
\sum_{i\,\delta}(\Delta_{i\,\delta}\langle\Delta_{i\,\delta}^\dagger\rangle 
+\Delta_{i\,\delta}^\dagger\langle\Delta_{i\,\delta}\rangle),
\label{one}
\end{equation}
where $c^\dagger_{i\,\sigma}$ creates an electron of spin $\sigma$ on the $i^{\, th}$
site and $t$ denotes an effective nearest-neighbor hopping. The $\langle
i\,j\rangle$ sum is over nearest-neighbor sites of a 2D square lattice,
and in the second term $\delta$ connects $i$ to its nearest-neighbor sites.
In Eq.~(\ref{one}) one
could, of course, add a next-near-neighbor hopping $t^\prime$ and a
chemical potential term.  Here, for simplicity and to refrain from
further approximations, we have set $t^\prime$ and
the chemical potential equal to zero \cite{comm}. We will assume that below a mean
field temperature $T_c^{MF}$, a $d_{x^2-y^2}$-gap amplitude 
forms with $\Delta \sim 2\, T^{MF}_c$.
The detailed temperature dependence of $\Delta$ is not central, as we
are not interested in the region around $T_c^{MF}$ where the pseudogap
closes. The
important point for our calculations is simply that a $d_{x^2-y^2}$-gap
amplitude of order $2T^{MF}_c$ in magnitude forms as $T$ drops
below $T^{MF}_c$ so that
\begin{equation}
\langle\Delta_{i\,\delta}^\dagger\rangle=
\frac{1}{\sqrt{2}}\langle c_{i\,\uparrow}^\dagger c_{i+\delta\,\downarrow}^\dagger
-c_{i\,\downarrow}^\dagger c_{i+\delta\,\uparrow}^\dagger\rangle=
\Delta\,e^{i \,\Phi_{i \delta}},\label{two1}
\end{equation}
with
\begin{equation}
\Phi_{i \delta}=\left\{\begin{array}{l@{\quad \mathrm{for} \quad}l}
(\varphi_i + \varphi_{i+\delta})/2 & \text{$\delta$ in x-direction,} \\
(\varphi_i + \varphi_{i+\delta})/2 +\pi & \text{$\delta$ in y-direction.} 
\end{array} \right. \label{two2}
\end{equation}
We then determine the fluctuating phases from a Monte Carlo
calculation using an effective 2D $XY$-free energy
\begin{equation}
F\left[\varphi_i\right] = - E_1 \sum_{\langle ij\rangle}
\cos\left(\varphi_i-\varphi_j\right)\ ,
\label{two}
\end{equation}
with $E_1$ adjusted to set the Kosterlitz-Thouless \cite{KTxx} transition
temperature $T_{KT}$ equal to some fraction of $T^{MF}_c$.
Specifically, for the present calculation we will set $T_{KT} \simeq
T^{MF}_c/5$. Here, we have the recent scanning tunneling results
\cite{Kug01} for $Bi_2Sr_2CuO_{6+\delta}$ in mind, where $T_c \simeq 10K$
and the pseudogap regime extends to 50 or 60K, which we take as
$T_c^{MF}$.

In principle, the $XY$ action, which
determines the fluctuations of the phases, arises from integrating out the
shorter wavelength fermion degrees of freedom including those responsible
for the local pair amplitude and the internal $d_{x^2-y^2}$ structure of
the pairs. In general this leads to a $\tau$-dependent quantum action
as well as a coupling energy $E_1$, whose temperature dependence
is determined by the many-body interactions
of the microscopic system.  There have been various discussions
regarding the regime over which a classical action is appropriate for the
cuprates \cite{Eme95,Par00,Ben01}. 
Here, however, we will proceed
phenomenologically using the classical action, Eq.~(\ref{two}), and
neglecting the temperature dependence of $E_1$.  Furthermore, we will
use the 2D form of Eq.~(\ref{two}). One knows that for the
layered cuprates there is a crossover from 2D to 3D $XY$ behavior near
$T_c$ \cite{Ref1}. Our point of view is that away from this crossover regime,
a 2D model is certainly suitable and on the finite size lattice that we will study,
the system becomes effectively ordered as $T$ approaches $T_{KT}$ and the
correlation length exceeds the lattice size. So $E_1$ will simply be
used to set $T_{KT}\equiv T_c$. A crucial physical point
that will be taken into account in our analysis is that the basic length
scale of the $\varphi$-field is larger than the Cooper-pair size $\xi_0$.
Thus, although this is a clearly simplified model, we believe that its
 solution provides useful insight into the
experimental consequences of the phase fluctuation pseudogap scenario.
It is the central aim of this paper to verify this by
comparison with the STM experiments and reproduction of some of their characteristic 
and salient features.

The calculation of the density of states for an $L\times L$ periodic
lattice now proceeds as follows \cite{Ref,Dag98}.  A set of phases $\{\varphi_i\}$ is
generated by a Monte Carlo (MC) importance sampling procedure, in which the
probability of a given configuration is proportional to $\exp
(-F[\varphi_i]/T)$ with $F$ given by Eq.~(\ref{two}).  With $\{\varphi_i\}$
given, the Hamiltonian of Eq.~(\ref{one}) is diagonalized and the
single particle density of states $N(\omega, T, \{\varphi_i\})$ is calculated.
Further MC $\{\varphi_i\}$ configurations are generated and an
average density of states $N(\omega, T)= \langle N(\omega, T,
\{\varphi_i\})\rangle$ at a given temperature is determined.  

As noted above, our point of view is that the $XY$ action, used in the MC
simulations, in principle arises from integrating out the shorter wavelength
fermion degrees of freedom up to the scale of the
Cooper-pair size, so that only the {\it center of mass} pair
 phase fluctuations
are important. Thus, the scale of the lattice spacing for $F[\varphi_i]$ is
set by the pair size coherence length $\xi_0 \sim v_F/\pi\Delta_0$ and is of
order 3 to 4 times the basic $Cu$-$Cu$ lattice spacing of the fermion
Hamiltonian Eq.~(\ref{one}).
Now the computationally intensive part of the calculation is the diagonalization
of $H$ and in order to get meaningful results as $T$ approaches $T_{KT}$, we found
it necessary to average over a large number of Monte Carlo $\{\varphi_i\}$
configurations. This requires that some compromise be made with
respect to the lattice size.  The results, we will present, are for a $32\times
32$ Hamiltonian lattice. However, if we were to take $\xi_0 \sim 4$ lattice
spacings, this would lead to only an $8\times 8$ lattice for the $\varphi_i$
simulations. This would not allow a sufficient range for the
Kosterlitz-Thouless phase coherence length to grow as $T$ approaches
$T_{KT}$. Thus, we have chosen to set $\Delta = 1.0t$ giving $\xi_0\sim 1$
so that the $\varphi_i$ simulation can be carried out on the same $L\times
L$ lattice that is used for the diagonalization of $H$. The important
physical point is that this procedure effectively cuts off phase
fluctuations on a scale less than the Cooper-pair size, $\xi_0$. Thus, the
phase coherence length is always larger than the Cooper-pair size when $T$
is less than $T^{MF}_c$. Consequently, our results differ from earlier work
\cite{Lam01}, which found that the pseudogap regime due to fluctuating phases 
extended
only about 20\% above $T_c$, in contrast to the Bi tunneling
experiments \cite{Ren98M,Kug01} and the recent Nernst-effect results
\cite{Wan02}.  In the work of Ref.~\cite{Lam01}, parameters were used which set
the basic scale of the phase correlation length to be much smaller than
$\xi_0$ and, therefore, the phase correlation length exceeded $\xi_0$ only in
a narrow temperature region set by a fraction of $T_{KT}$. We believe that 
this is not the correct phenomenology.

\begin{figure}[h]
\begin{center}
\epsfig{file=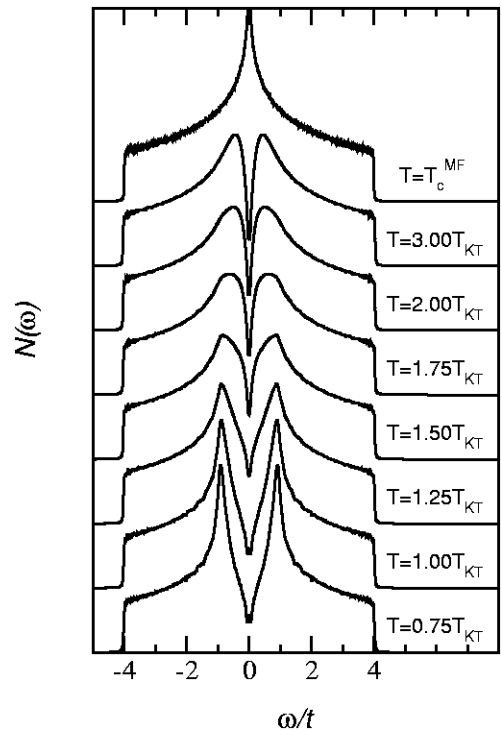,width=6.5cm}
\end{center}
\caption[]{Single particle density of states $N(\omega)$ for different temperatures
$T$ for a $32 \times 32 $ lattice with $\Delta=1.0t$ and $T_{KT}=0.1t$. A pseudogap appears
below $T_c^{MF} \simeq 0.5t$ and coherence peaks develop as $T$ approaches $T_{KT}$.}
\label{n_omega}
\end{figure}

Results for $N(\omega, T)$  are
shown in Fig.~\ref{n_omega}.
For each temperature we have generated up to $25,000$ independent MC $\{\varphi_i\}$
configurations, diagonalized $H$ for each of these configurations, and computed $\langle N(\omega, T, \{\varphi_i\}) \rangle$.
In these calculations, as discussed above, we have set $\Delta=1.0t$ corresponding
to $T^{MF}_c \simeq 0.5t$ and selected $E_1$ so that $T_{KT}=0.1t$ \cite{Tkt}. 
In order to reduce finite-size effects, we employ a very effective scheme recently suggested
by F.~F.~Assaad \cite{assaad}.

For $T>T^{MF}_c$, the gap amplitude vanishes and
the density of states exhibits the usual Van Hove peak at $\omega=0$. For
$T<T^{MF}_c$, the presence of a finite gap amplitude  gives
rise to a pseudogap whose size is set by $2 \Delta$.   Then, as $T$
approaches $T_{KT}$ and the $XY$ phase correlation length rapidly increases,
coherence peaks evolve, the separation of which is determined by
$2\Delta$.  An important point is that the scale 
in temperature over which the 
evolution of the coherence peaks occurs, is set by some fraction of $T_{KT}$ which means that it appears 
suddenly on a scale set by $T^{MF}_c$.

\begin{figure}[h]
\begin{center}
\epsfig{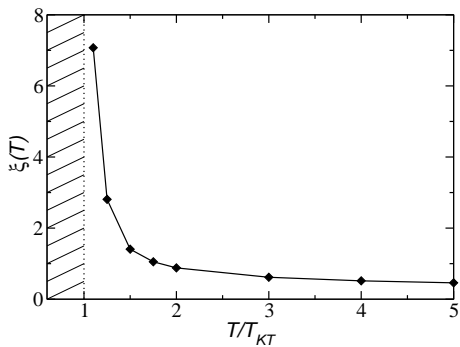}
\end{center}
\caption[]{The {\sl effective} correlation length $\xi(T)$ versus $T/T_{KT}$
for the $32 \times 32$ lattice. Here $T_c^{MF}/T_{KT} \simeq 5$ so that the pseudogap regime which
extends from $T/T_{KT} \simeq 1.5$ to $5$ is large compared to the superconducting region which
extends from $0$ to $T/T_{KT}=1$. The pronounced increase of $\xi(T)$ occurs over a narrow temperature
region, on a scale set by $T_c^{MF}$, as $T_{KT}$ is approached.}
\label{korrel}
\end{figure}

An {\sl effective} correlation length $\xi(T)$, extracted by fitting 
an exponential form to the correlation function
\begin{equation}
C(\ell) = \left\langle
e^{-i\varphi_{i+\ell}}e^{i\varphi_i}\right\rangle
\label{three}
\end{equation}
is plotted versus $T$ in Fig.~\ref{korrel} for our $32 \times 32$ lattice.
The rapid onset of $\xi(T)$ as $T_{KT}$ is approached 
 is clearly seen.  It is this sudden increase of $\xi(T)$ that is
responsible for the appearance of the coherence peaks as $T$ approaches
$T_{KT}$.
This effect is further enhanced by the $2D$ to $3D$ crossover that occurs
in the actual materials.

\begin{figure}[h]
\begin{center}
\epsfig{file=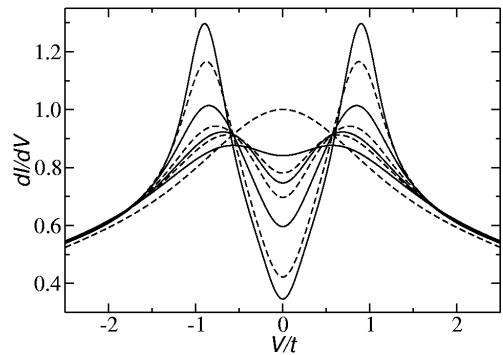,width=6.5cm}
\end{center}
\caption[]{Tunneling conductance, $\frac{dI}{dV}$, normalized to its value at $T_c^{MF}$ and $V=0$, for different 
temperatures. Solid curves are for
$T=\{0.75,1.25,1.75,3.00\}\,\,T_{KT}$, dashed curves for $T=\{1.00,1.50,2.00\}\,\,T_{KT}$ and $T_c^{MF}$ 
($\frac{dI}{dV}|_{V=0}$ is increasing with $T$).}
\label{didv}
\end{figure}

\begin{figure}[h]
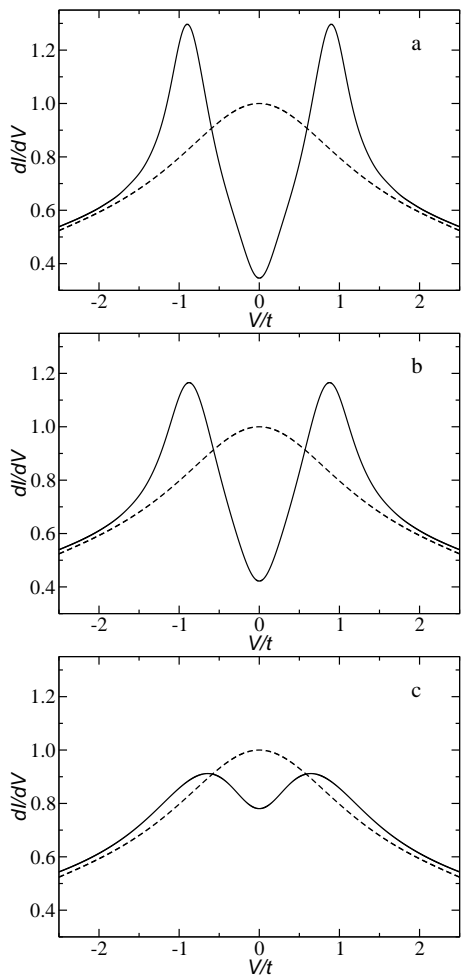

\begin{center}
\epsfig{file=fig4a.eps,width=6cm}\\
\epsfig{file=fig4b.eps,width=6cm}\\
\epsfig{file=fig4c.eps,width=6cm}
\end{center}
\caption[]{Temperature dependence of $\frac{dI}{dV}$ normalized to its value at $T_c^{MF}$ and $V=0$. 
The solid curves are for $T=0.75 T_{KT}$ (a),
$T= T_{KT}$ (b) and $T=2 T_{KT}$ (c). The dashed curve in all three figures is for 
$T=T_c^{MF}\simeq 5 T_{KT}$.}
\label{didv_x}
\end{figure}

In order to compare these results for $N(\omega, T)$ with scanning
tunneling spectra $dI/dV$, we have calculated $dI(V,T)/dV$ using the standard
quasi-particle expression for the tunneling current,
\begin{equation}
\frac{dI(V,T)}{dV}\propto\int N(\omega)\,\frac{\partial f(\omega-V)}{\partial V} \,\,d\omega.
\label{five}
\end{equation}
Here, $f(\omega) = (\exp(\omega/T)+1)^{-1}$ is the usual 
Fermi factor.
Results for $dI(V,T)/dV$ are displayed in Fig.~\ref{didv}.  
The effect of the Fermi
factors is to provide a thermal smoothing of the quasi-particle density of
states over a region of order $2T$.  This becomes significant at the higher
temperatures and the prominent pseudogap dependence of $N(\omega, T)$ seen
in Fig.~\ref{n_omega} is
smoothed out in $dI/dV$.  In Fig.~\ref{didv_x}, $dI/dV$ results are shown as  solid
curves for $T=0.75T_{KT}$ (Fig.~\ref{didv_x}a),  $T=T_{KT}$ (Fig.~\ref{didv_x}b) and 
$T=2T_{KT}$ (Fig.~\ref{didv_x}c). 
The dashed curve is for
$T=T^{MF}_c\simeq 5 T_{KT}$. One sees that the size of the pseudogap scales with the
spacing between the coherence peaks and evolves continuously out of the
superconducting state. The pseudogap persists over a large temperature
range measured in units of $T_{KT}$, becoming smoothed out by the thermal
effects as $T$ approaches
$T^{MF}_c$ and vanishing above $T^{MF}_c$.

Our numerical results for $dI(V,T)/dV$ are similar to recent scanning tunneling measurements of Bi2212 and Bi2201
\cite{Ren98M,Kug01}.  Also in these experiments the superconducting gap for $T<T_{KT}$ evolves continuously into the
pseudogap regime, which extends up to $T=T_c^{MF}$. The coherence peaks appear suddenly as
$T_{KT}$ is approached. At higher temperatures, the pseudogap fills in rather than closing 
and the temperature range associated with the pseudogap regime can be large
compared with the size of the superconducting regime.

Summarizing, in order to develop a more quantitative understanding for the role of phase fluctuations, 
we have provided a numerical solution of a simplified model which,
nevertheless, contains the key ideas of the cuprate phase fluctuation
pseudogap scenario. Here the center of mass pair-phase
fluctuations of a BCS $d$-wave model were determined from a classical 2D
 $XY$ action by
means of a Monte Carlo simulation. 
The resulting tunneling conductance ($dI/dV$)
reproduces characteristic and salient features of recent STM studies of Bi2212 and
Bi2201 suggesting that the pseudogap behavior observed in these experiments
 arises from phase fluctuations of the $d_{x2-y2}$-pairing gap.

We would like to acknowledge useful discussions with S.~A.~Kivelson and A.~Paramekanti.
This work was supported by the DFG under Grant No.~Ha 1537/16-2 and
AR 324/3-1, by the Bavaria California Technology Center (BaCaTeC),
the superconducting project KONWHIR OOPCV and by the U.~S.~Department of
Energy under Grant No.~DE-FG03-85ER45197.
The calculations were carried out at the high-performance computing centers 
HLRS (Stuttgart) and LRZ (M\"unchen).


\begin{references}
\bibitem{comm} We have checked to some extent that our results do not depend appreciably on $t^{\prime}$ or $\mu$.
\end{references}
\end{document}